\documentclass[10pt,conference]{IEEEtran}
\IEEEoverridecommandlockouts
\usepackage{cite}
\usepackage{amsmath,amssymb,amsfonts}
\usepackage{algorithmic}
\usepackage{graphicx}
\usepackage{textcomp}
\usepackage{balance}
\usepackage{xcolor}
\def\BibTeX{{\rm B\kern-.05em{\sc i\kern-.025em b}\kern-.08em
    T\kern-.1667em\lower.7ex\hbox{E}\kern-.125emX}}
    
\usepackage{xspace}
\usepackage{fp}
\usepackage[autolanguage]{numprint}
\newcommand*\np[2][z]{
\ifx z#1%
$\numprint{#2}$%
\else%
$\numprint[#1]{#2}$%
\fi\xspace
}
\usepackage{adjustbox}

\usepackage{booktabs}
\usepackage{multirow}
\usepackage{threeparttable}

\usepackage[bookmarks=false]{hyperref}
\usepackage{tcolorbox,tikz}
\usepackage{pgfplots}
\usepackage{pgfplotstable}
\usepackage{filecontents}
\usepgfplotslibrary{statistics}
\pgfplotsset{compat=1.14} 
\usepackage{silence}
\usepackage{listings,xcolor,beramono}
\usepackage{tikz}

\makeatletter
\newenvironment{btHighlight}[1][]
{\begingroup\tikzset{bt@Highlight@par/.style={#1}}\begin{lrbox}{\@tempboxa}}
{\end{lrbox}\bt@HL@box[bt@Highlight@par]{\@tempboxa}\endgroup}

\newcommand\btHL[1][]{%
  \begin{btHighlight}[#1]\bgroup\aftergroup\bt@HL@endenv%
}
\def\bt@HL@endenv{%
  \end{btHighlight}%
  \egroup
}
\newcommand{\bt@HL@box}[2][]{%
  \tikz[#1]{%
    \pgfpathrectangle{\pgfpoint{1pt}{0pt}}{\pgfpoint{\wd #2}{\ht #2}}%
    \pgfusepath{use as bounding box}%
    \node[anchor=base west, fill=red!20,outer sep=0pt,inner xsep=1pt, inner ysep=0pt, rounded corners=1pt, minimum height=\ht\strutbox+.8pt,#1]{\raisebox{.8pt}{\strut}\strut\usebox{#2}};
  }%
}
\makeatother

\newcommand*{\boxplott}[8]{%
  \addplot+[
    mark=star,
    line width=.1mm, 
    fill=black!20, 
    draw=black!60,
    boxplot prepared={
      lower whisker={#7},
      lower quartile={#4},
      median={#3},
      upper quartile={#5},
      upper whisker={#6},
    },
  ] coordinates{#8};
}

\definecolor{blau_1a}{RGB}{93,133,195}
\definecolor{blau_2a}{RGB}{0,156,218}
\definecolor{gruen_3a}{RGB}{80,182,149}
\definecolor{gruen_4a}{RGB}{175,204,80}
\definecolor{gruen_5a}{RGB}{221,223,72}
\definecolor{orange_6a}{RGB}{255,224,92}
\definecolor{orange_7a}{RGB}{248,186,60}
\definecolor{rot_8a}{RGB}{238,122,52}
\definecolor{rot_9a}{RGB}{233,80,62}
\definecolor{lila_10a}{RGB}{201,48,142}
\definecolor{lila_11a}{RGB}{128,69,151}

\definecolor{blau_1b}{RGB}{0,90,169}
\definecolor{blau_2b}{RGB}{0,131,204}
\definecolor{gruen_3b}{RGB}{0,157,129}
\definecolor{gruen_4b}{RGB}{153,192,0}
\definecolor{gruen_5b}{RGB}{201,212,0}
\definecolor{orange_6b}{RGB}{253,202,0}
\definecolor{orange_7b}{RGB}{245,163,0}
\definecolor{rot_8b}{RGB}{236,101,0}
\definecolor{rot_9b}{RGB}{230,0,26}
\definecolor{lila_10b}{RGB}{166,0,132}
\definecolor{lila_11b}{RGB}{114,16,133}

\usepackage{inconsolata}
\definecolor{pblue}{rgb}{0.13,0.13,1}
\definecolor{pgreen}{rgb}{0,0.5,0}
\definecolor{pred}{rgb}{0.9,0,0}

\definecolor{diff_add}{HTML}{D1FFCF}
\definecolor{diff_remove}{HTML}{FF9C96}

\usepackage[autostyle]{csquotes}
    
\newcommand\numberToBeChecked[1]{{#1}}

\newcommand{\nbPatchesWithClonesNum}{2064}
\newcommand{\nbSingleCloneChangePatches}{1452}
\newcommand{\nbAnalyzedPatches}{3049}

\newcommand{\nbPatchesMultiHunkNum}{3475}

\newcommand{\nbAnalyzedPatchesToPrint}{\numberToBeChecked{\np{\nbAnalyzedPatches}}\xspace}
\newcommand{\nbProjects}{\numberToBeChecked{\np{96}}\xspace}

\newcommand{\nbSingleChanges}{\numberToBeChecked{\np{25539}}\xspace}
\newcommand{\nbPatches}{\numberToBeChecked{\np{11624}}\xspace}
\newcommand{\nbPatchesAfterRemovalDuplicates}{\numberToBeChecked{\np{10290}}\xspace}
\newcommand{\nbPatchesMultiHunk}{\numberToBeChecked{\np{\nbPatchesMultiHunkNum}}\xspace}
\newcommand{\nbPatchesWithClones}{\numberToBeChecked{\np{\nbPatchesWithClonesNum}}\xspace}
    
\newcommand{\ShowAbsoluteNumber}[1]{%
\ifnum #1<10%
{\hspace*{0pt}#1}%
\else%
\ifnum #1<100%
{\hspace*{0pt}#1}%
\else%
\ifnum #1<1000%
{\hspace*{0pt}#1}%
\else%
{\numprint{#1}}%
\fi%
\fi%
\fi%
}

\newcommand{\ShowPercentage}[2]{%
\FPeval\percent{round(#1/#2*10000,0)}%
\FPeval\percentageOneDecimal{round(#1/#2*100,1)}%
\FPeval\percentageNoDecimal{round(#1/#2*100,0)}%
\ifnum \percent=0%
    {\np[\%]{0}}%
\else%
    \ifnum \percent<10%
        {$<$\np[\%]{0.1}}%
    \else%
    \ifnum \percent<1000%
        {\np[\%]{\FPprint{percentageOneDecimal}}}%
    \else%
        {\np[\%]{\FPprint{percentageNoDecimal}}}%
    \fi\fi%
\fi%
\xspace
}
\newlength\BARSIZE  
\newcommand{\inlinechart}[2]{%
\FPeval{\BLACKBARSIZE}{#1/#2}\textcolor{black!80}{\rule{\BLACKBARSIZE\BARSIZE}{1.6ex}}%
\FPeval{\BLACKBARSIZE}{1 - (#1/#2)}\textcolor{black!10}{\rule{\BLACKBARSIZE\BARSIZE}{1.6ex}}%
}

\newcommand*\percentage[3][v]{%
\ifx e#1%
    \ShowPercentage{#2}{#3}\else%
\ifx d#1%
    \ShowPercentage{#2}{#3} (\np{#2}/\np{#3})\else%
\ifx q#1%
    \np{#2}/\np{#3}(\ShowPercentage{#2}{#3})\else%
\ifx p#1%
    \np{#2}(\ShowPercentage{#2}{#3})\else%
\ifx c#1%
    \inlinechart{#2}{#3}%
\else%
    \np{#2}%
    \ifx r#1%
        /\np{#3}%
    \fi%
    \hspace*{0.5ex}(\ShowPercentage{#2}{#3}) %
    \inlinechart{#2}{#3}%
    \xspace
\fi\fi\fi\fi\fi%
}

\IEEEaftertitletext{\vspace{-1.4\baselineskip}}

\begin{document}

\title{A large-scale study on human-cloned changes for automated program repair}

\author{\IEEEauthorblockN{Fernanda Madeiral}
\IEEEauthorblockA{KTH Royal Institute of Technology \\
Stockholm, Sweden \\
fer.madeiral@gmail.com}
\and
\IEEEauthorblockN{Thomas Durieux}
\IEEEauthorblockA{KTH Royal Institute of Technology \\
Stockholm, Sweden \\
thomas@durieux.me}
}


\maketitle

\begin{abstract}
Research in automatic program repair has shown that real bugs can be automatically fixed. However, there are several challenges involved in such a task that are not yet fully addressed. As an example, consider that a test-suite-based repair tool performs a change in a program to fix a bug spotted by a failing test case, but then the same or another test case fails. This could mean that the change is a partial fix for the bug or that another bug was manifested. However, the repair tool discards the change and possibly performs other repair attempts. One might wonder if the applied change should be also applied in other locations in the program so that the bug is fully fixed. In this paper, we are interested in investigating the extent of bug fix changes being cloned by developers within patches. Our goal is to investigate the need of multi-location repair by using identical or similar changes in identical or similar contexts. To do so, we analyzed \nbAnalyzedPatchesToPrint multi-hunk patches from the ManySStuBs4J dataset, which is a large dataset of single statement bug fix changes. We found out that \numberToBeChecked{\percentage[e]{\nbPatchesWithClonesNum}{\nbAnalyzedPatches}} of the multi-hunk patches contain at least one change clone group. Moreover, most of these patches (\numberToBeChecked{\percentage[e]{\nbSingleCloneChangePatches}{\nbPatchesWithClonesNum}}) are strictly-cloned ones, which are patches fully composed of changes belonging to one single change clone group. Finally, most of the strictly-cloned patches (\numberToBeChecked{\percentage[e]{1286}{\nbSingleCloneChangePatches}}) contain change clones with identical changes, independently of their contexts. We conclude that automated solutions for creating patches composed of identical or similar changes can be useful for fixing bugs.
\end{abstract}


\section{Introduction}

Fixing the source code of software systems is an inherent activity of software developers' jobs. Research has been conducted for decades on automated solutions, e.g., software testing, in order to support developers in the process of finding, understanding, and fixing bugs. A more ambitious desired solution is the \textit{automatic repair} of the source code \cite{Monperrus2018bibliography}, which has been extensively explored by the research community in the last decade, and research has shown that automatic repair tools do fix real bugs automatically.

There are, however, several challenges not yet fully-addressed. As an example, consider a test-suite-based repair tool. Such a tool receives a program and at least one failing test case as the specification of the existing undesired behavior of the program. Then, the tool performs a change in the program to fix the bug, but the patched program still causes test failures. The repair tool discards the change and possibly performs other repair attempts. However, the change could be a partial fix for the bug, or the correct fix for the originally-exposed bug but then another one was manifested.

The described scenario stresses the need for multi-location bug repair, which currently has been explored by only a few works (e.g., \cite{Mechtaev2016,Saha2019}). Bugs that require changes in multiple locations are hard to fix.
A way to make research progress is to first focus on simple cases, such as applying identical or similar changes in different locations, i.e., cloning the changes.

\vspace{5pt}
\noindent\textit{Problem statement.} There has been research relating code clones and bugs (e.g., \cite{Steidl2013,Mondal2019}), but, to the best of our knowledge, there is no study on cloned bug fix changes at scale and a classification of them based on the similarity of changes and their contexts.
\vspace{5pt}

In this paper, we attempt to shed light on the extent to what developers apply \textit{change clones} for fixing bugs.
To do so, we used the ManySStuBs4J dataset \cite{Karampatsis2020ManySStuBs4J}, which contains single statement bug fix changes from \nbPatchesAfterRemovalDuplicates patches from \nbProjects projects. We manually analyzed \nbAnalyzedPatchesToPrint multi-hunk patches, i.e., patches composed of changes in multiple non-contiguous locations, and annotated them with information on change clones that were committed together. Moreover, we classified the change clone groups in \numberToBeChecked{five} types of change clones that we define in this paper.

We found out that change clones are frequently present in developer patches, appearing in \numberToBeChecked{\percentage[d]{\nbPatchesWithClonesNum}{\nbAnalyzedPatches}} of the analyzed multi-hunk patches. Moreover, \numberToBeChecked{\percentage[d]{\nbSingleCloneChangePatches}{\nbPatchesWithClonesNum}} of them are \textit{strictly-cloned patches}, which are fully composed of changes belonging to one single change clone group. Finally, \numberToBeChecked{\percentage[d]{1286}{\nbSingleCloneChangePatches}} of the strictly-cloned patches contain change clones with identical changes regardless of their contexts.

\vspace{5pt}
\noindent\textit{Contribution.} Our contribution is a large-scale study on human-cloned changes. We conclude that automated solutions for creating patches composed of identical or similar changes in different locations of programs can be useful for fixing software bugs.

\vspace{5pt}
\noindent\textit{Data availability.} The data produced in this study is publicly available at
\url{https://github.com/software-bugs/change-clone}.

\section{Method}

\subsection{Definitions}\label{sec:definitions}

The first step of our study was to define \textit{change clones} and \textit{change clone types}. By consulting the literature, we found out that \textit{code clones} and the well-known \textit{code clone types} are closely related to our idea of \textit{change clones} and \textit{change clone types}. However, the existing definitions on code clones cannot be directly reused due to the difference of nature between code clone and change clone.
In this section, we motivate the need for new definitions for change clones and their types, and then we present the actual definitions supported by examples.

First, consider the definition of the code clone types provided by Roy and Cordy \cite{Roy2007}:

\vspace{1pt}
\blockquote{
Type I: Identical code fragments except for variations in whitespace (may be also variations in layout) and comments.

Type II: Structurally/syntactically identical fragments except for variations in identifiers, literals, types, layout, and comments.

Type III: Copied fragments with further modifications. Statements can be changed, added, or removed in addition to variations in identifiers, literals, types, layout, and comments.

Type IV: Two or more code fragments that perform the same computation but implemented through different syntactic variants.}
\vspace{1pt}

Then, consider the patch presented in \autoref{code:typeB}, which contains two single statement changes. The calls to the method \texttt{Long.valueOf} were removed from both statements while their arguments were kept in the code. The changes themselves are the same in both statements, but their contexts albeit similar are different. Those changed lines are change clones. However, we cannot classify them as the traditional code clones of Type I, because the contexts of the changes are different. We also cannot simply classify the changed lines as code clones of Type II, because the changes are identical.

\vspace{3pt}
\begin{lstlisting}[caption={Identical changes, similar contexts (Type B).},label={code:typeB}]
- assigneeNode.put("id", `Long.valueOf(`userTask.getAssignee()`));`
+ assigneeNode.put("id", userTask.getAssignee()@);@
...
- candidateUserNode.put("id", `Long.valueOf(`candidateUser`));`
+ candidateUserNode.put("id", candidateUser@);@
\end{lstlisting}

\textit{Code clones} are identical or similar code fragments that already exist in the source code. However, \textit{change clones} can be identical or similar in two aspects: the actual changes and the contexts where the changes were applied. For this reason, we define change clones and their types in this paper, taking into account the types of code clones and the two aspects of similarity in change clones, that is, the actual changes and their contexts.

\vspace{5pt}
\noindent\textbf{Definition 1: Change clone.} A change clone is a source code change that is identical or similar to another change. There are two aspects of similarity between two changes: the actual changes and their contexts.

\vspace{5pt}
\noindent\textbf{Definition 2: Change clone type.} A change clone type specifies how change clones are similar.

\break

\begin{lstlisting}[caption={Identical changes, identical contexts (Type A).},label={code:typeA}]
- s `+=` s.length() + endString + s;
+ s  @=@ s.length() + endString + s;
...
- s `+=` s.length() + endString + s;
+ s  @=@ s.length() + endString + s;
\end{lstlisting}

\vspace{-3pt}

\begin{lstlisting}[caption={Similar changes, identical contexts (Type C).},label={code:typeC}]
- databaseFormatter = new `DatabaseFormatterOracle`();
+ databaseFormatter = new @DatabaseFormatterDb2@();
...
- databaseFormatter = new `DatabaseFormatterOracle`();
+ databaseFormatter = new @DatabaseFormatterPostgres@();
\end{lstlisting}

\vspace{-3pt}

\begin{lstlisting}[caption={Similar changes, similar contexts (Type D).},label={code:typeD}]
- callsPerKey *= `numKey1 /` (double) numRecords1;            
+ callsPerKey *= (double)numRecords1 @/ numKey1@;
...
- callsPerKey *= `numKey2 /` (double) numRecords2;            
+ callsPerKey *= (double) numRecords2 @/ numKey2@;
\end{lstlisting}

\vspace{-3pt}

\begin{lstlisting}[caption={Identical changes, not similar contexts (Type E).},label={code:typeE}]
- response.write(data + `"<||>"`);
+ response.write(data + @END@);
...
- return message + `"<||>"`;
+ return message + @END@;
\end{lstlisting}

\vspace{4pt}
We defined \numberToBeChecked{five} change clone types, as explained as follows.

\vspace{5pt}
\noindent\textbf{Type A}: The changes are identical and their contexts are also identical, as shown in \autoref{code:typeA}.

\vspace{3pt}
\noindent\textbf{Type B}: The changes are identical, and their contexts are structurally similar, possibly containing variations in identifiers, literals, types, layout, comments, and added or removed code elements, as shown in \autoref{code:typeB}.

\vspace{3pt}
\noindent\textbf{Type C}: The changes are structurally similar, possibly containing variations in identifiers, literals, types, layout and comments, and their contexts are identical, as shown in \autoref{code:typeC}.

\vspace{3pt}
\noindent\textbf{Type D}: The changes and their contexts are structurally similar, possibly containing variations in identifiers, literals, types, layout, and comments, as shown in \autoref{code:typeD}.

\vspace{3pt}
\noindent\textbf{Type E}: The changes are identical, and their contexts are not structurally similar, as shown in \autoref{code:typeE}.

\vspace{5pt}
Note that our definitions of change clone types are inspired by the code clone types.
\autoref{tab:change_clone_types} presents the relation between them.
The Types A, B, C, and D are directly supported by the definitions of Types I, II, and III. Type E resembles the copy and paste operations of fine-grained code elements.
We do not take into account Type IV of code clones because it involves code semantic analysis, which is out of the scope of this work due to its scale.

\begin{table}[b]
    \vspace{-12pt}
    \centering
    \caption{Relation between change clone types and code clone types.}
    \label{tab:change_clone_types}
    \begin{tabular}{@{}l|ccccc@{}}
        \toprule
                & Type A & Type B              & Type C  & Type D  & Type E \\
        \midrule
        Change  & Type I & Type I              & Type II & Type II & Type I \\
        Context & Type I & Type II \& Type III & Type I  & Type II & --     \\
        \bottomrule
    \end{tabular}
\end{table}

\vspace{5pt}
\noindent\textbf{Definition 3: Change clone group.} A change clone group is a set of two or more identical or similar change clones.

\subsection{Data collection}

Our work is an initial effort towards understanding change clone types in patches. We aim to do so at large scale, and for that we need a large data source. There are datasets that could be used in our study, such as Defects4J \cite{Just2014Defects4J} and Bears \cite{Madeiral2019Bears}, which contain bugs and their respective patches. Those datasets are curated and focused on providing researchers with reproducible bugs. Naturally, they are not large (by large, we mean a dataset with hundreds of data entries), so they are not the ideal fit for our study. Recently, researchers created the ManySStuBs4J dataset \cite{Karampatsis2020ManySStuBs4J}, which does not necessarily contain reproducible bugs, but it is large in terms of bug fix changes. Therefore, we use this dataset in our study.
Note that there are two versions of the ManySStuBs4J dataset: a small and a large one. Our analysis (further explained in \autoref{sec:data-analysis}) is manual, therefore we use the small ManySStuBs4J dataset, which contains \nbSingleChanges single statement changes.

\subsection{Data preprocessing}

The ManySStuBs4J dataset is organized at the level of single statement changes. Our work is conducted at the level of patches, since we want to investigate change clones within patches. Therefore, we first grouped the changes by commit (patch). We found out that the \nbSingleChanges single statement changes are from \nbPatches patches.
Then, we removed patches with duplicate diffs. This process resulted in \nbPatchesAfterRemovalDuplicates patches.

\subsection{Data analysis}\label{sec:data-analysis}

We manually analyzed the patches and annotated the single statement changes that are change clones with their types. The actual analyzed patches were selected as follows. First, we discarded single-hunk patches, i.e., patches composed of changes in a single contiguous location, resulting in \nbPatchesMultiHunk multi-hunk patches. We then analyzed \nbAnalyzedPatchesToPrint patches that have at maximum \numberToBeChecked{six} changes to keep the manual effort reasonable.

\section{Results}\label{sec:results}

We first found out that \numberToBeChecked{\percentage[d]{\nbPatchesWithClonesNum}{\nbAnalyzedPatches}} of the analyzed multi-hunk patches contain at least one change clone group. Surprisingly, \numberToBeChecked{\percentage[d]{\nbSingleCloneChangePatches}{\nbPatchesWithClonesNum}} of these patches are \textit{strictly-cloned patches}, which are fully-composed of changes belonging to one single change clone group. This means that a repair tool could generate these patches by only applying the same or similar changes in the same or similar contexts.

\autoref{fig:frequency-clone-types} shows the frequency of the \nbPatchesWithClones patches that contain at least one change clone group, per change clone type. Change clones of Types A and B are the most frequent ones by far. Change clones with identical changes regardless of their contexts (change clones of Types A, B, and E) are present in \numberToBeChecked{\percentage[d]{1286}{\nbSingleCloneChangePatches}} of strictly-cloned patches.

\pgfplotstableread[col sep=comma, header=true]{frequency-clone-types.csv}{\datatable}
\pgfplotstablegetrowsof{\datatable}
\edef\numberofrows{\pgfplotsretval}

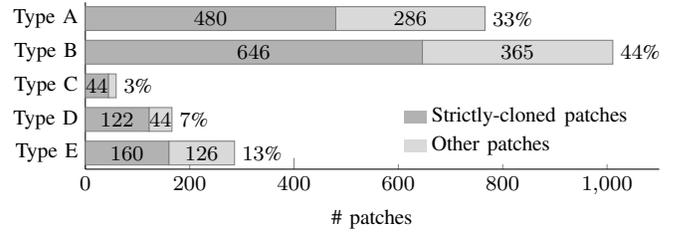
\begin{figure}[t]
  \centering
  \footnotesize
  \begin{tikzpicture}
  \pgfplotsset{
        show sum on top/.style={
            /pgfplots/scatter/@post marker code/.append code={%
                \node[
                    at={(normalized axis cs:%
                            \pgfkeysvalueof{/data point/x},%
                            \pgfkeysvalueof{/data point/y})%
                    },
                    anchor=west,
                ]
                {\pgfkeys{/pgf/fpu=true}\pgfmathparse{\pgfkeysvalueof{/data point/x}/2288*100}\pgfmathprintnumber[fixed,precision=0]{\pgfmathresult}\%};
            },
        },
    }
    \begin{axis}
    [xbar stacked,
    width=1.04\columnwidth,
    height=3.8cm,
    bar width=9pt,
	xlabel=\# patches,
    xlabel near ticks,
    axis x line* = bottom,
    axis y line* = left,
    xmin=0,
    xmax=1100,
    xtick={0,200,...,1000},
    ytick=data,
    yticklabels from table={\datatable}{type},
    yticklabel style={align=left},
    visualization depends on={meta < 20 \as \valueissmall},
    nodes near coords={\ifdim\valueissmall pt=1 pt \else \pgfmathprintnumber[precision=0]\pgfplotspointmeta \fi},
    every node near coord/.append style={
        anchor={center}
    },
    y dir=reverse,
    enlarge y limits=0.12,
    xtick pos=left,
    ytick pos=left,
    legend pos=south east,
    legend cell align={left},
    legend image code/.code={
        \draw [#1, draw=none] (0cm,-0.1cm) rectangle (0.3cm,0.07cm);
    },
    legend style={draw=none}
    ]
    
    \addplot[bar shift=0pt, draw=black!50, fill=black!30, point meta=explicit] table[x=unique, y expr=\coordindex, meta index=1]{\datatable};
    
    \addplot[bar shift=0pt, draw=black!50, fill=black!15, point meta=explicit,show sum on top] table[x=total, y expr=\coordindex, meta index=2]{\datatable};
    
    \legend{Strictly-cloned patches,Other patches}
    \end{axis}
  \end{tikzpicture}
  \vspace{-15pt}
  \caption{Frequency of patches per change clone type.}
  \label{fig:frequency-clone-types}
\end{figure}

\autoref{fig:distribution} shows the distribution of number of changes per change clone type, only considering the \numberToBeChecked{\np{\nbSingleCloneChangePatches}} strictly-cloned patches. The distributions are similar to each other, showing that usually the change clone groups are of size \numberToBeChecked{two}. Change clones of Type B are the only ones that also frequently happen in \numberToBeChecked{triples} and \numberToBeChecked{quadruples}. In some exceptional cases, the size of change clone groups are up to \numberToBeChecked{six}.
\autoref{fig:distribution_file} shows the distribution of number of changed files in strictly-cloned patches per change clone type. Change clone groups are usually in \numberToBeChecked{one} file, which is the median of the distributions, but they also frequently happen in \numberToBeChecked{two} and \numberToBeChecked{three} files for some types of change clones, i.e., Types A, B, and E.

\begin{figure}[t]
\footnotesize%
\adjustbox{valign=t}{\begin{minipage}[r][][c]{.233\textwidth}%
    \begin{tikzpicture}
    \begin{axis}[
        boxplot/draw direction=x,
        xmin=1.5,
        ymin=0.5,
        ymax=5.5,
        width=\linewidth,
        height=3.8cm,
        axis x line* = bottom,
        axis y line* = left,
        xlabel= \# changes,
        boxplot={
            box extend=0.5
        },
        y dir=reverse,
        boxplot/draw/median/.code={%
          \draw[mark size=2pt,/pgfplots/boxplot/every median/.try]
            \pgfextra
            \pgftransformshift{
              \pgfplotsboxplotpointabbox
                {\pgfplotsboxplotvalue{median}}
                {0.5}
            }
            \pgfsetfillcolor{white}
            \pgfuseplotmark{*}
            \endpgfextra
          ;
        },
        xmajorgrids = true,
        xtick style={draw=none}, 
        ytick style={draw=none}, 
        ytick={0.5,...,5.5},
        xtick={2,4,6},
        y tick label as interval,
        yticklabels={Type A,Type B,Type C,Type D,Type E}
    ]
    
\boxplott{0}{2.41875}{2.0}{2.0}{2.0}{2.0}{2.0}{ (0,3) (0,3) (0,3) (0,3) (0,6) (0,3) (0,6) (0,4) (0,3) (0,4) (0,3) (0,3) (0,6) (0,4) (0,4) (0,4) (0,3) (0,6) (0,3) (0,3) (0,4) (0,4) (0,3) (0,4) (0,3) (0,4) (0,3) (0,3) (0,5) (0,5) (0,5) (0,3) (0,3) (0,3) (0,3) (0,3) (0,4) (0,5) (0,3) (0,6) (0,3) (0,3) (0,3) (0,3) (0,3) (0,3) (0,3) (0,3) (0,4) (0,3) (0,4) (0,4) (0,4) (0,3) (0,4) (0,5) (0,3) (0,3) (0,4) (0,4) (0,6) (0,3) (0,3) (0,3) (0,5) (0,3) (0,5) (0,4) (0,4) (0,3) (0,3) (0,3) (0,4) (0,4) (0,3) (0,6) (0,4) (0,3) (0,3) (0,4) (0,4) (0,3) (0,4) (0,3) (0,3) (0,4) (0,4) (0,4) (0,3) (0,6) (0,4) (0,4) (0,3) (0,3) (0,3) (0,3) (0,5) (0,6) (0,6) (0,3) (0,5) (0,3) (0,3) (0,3) (0,5) (0,5) (0,6) (0,4) (0,4) (0,3) (0,6)}
\boxplott{0}{2.496904024767802}{2.0}{2.0}{3.0}{4.0}{2.0}{ (0,6) (0,5) (0,6) (0,5) (0,5) (0,5) (0,5) (0,5) (0,6) (0,5) (0,5) (0,5) (0,5) (0,5) (0,5) (0,6) (0,6) (0,6) (0,6) (0,5) (0,5) (0,5) (0,5) (0,5) (0,6) (0,6) (0,5) (0,5) (0,5) (0,5) (0,6) (0,6) (0,6) (0,6) (0,5) (0,5) (0,5) (0,6) (0,6) (0,5) (0,6) (0,5)}
\boxplott{0}{2.2954545454545454}{2.0}{2.0}{2.25}{2.25}{2.0}{ (0,3) (0,4) (0,3) (0,4) (0,3) (0,3) (0,3) (0,3) (0,3) (0,3) (0,3)}
\boxplott{0}{2.2868852459016393}{2.0}{2.0}{2.0}{2.0}{2.0}{ (0,3) (0,3) (0,3) (0,6) (0,3) (0,6) (0,3) (0,4) (0,4) (0,4) (0,4) (0,6) (0,4) (0,3) (0,3) (0,5) (0,5)}
\boxplott{0}{2.2375}{2.0}{2.0}{2.0}{2.0}{2.0}{ (0,3) (0,3) (0,4) (0,3) (0,3) (0,3) (0,4) (0,3) (0,4) (0,3) (0,3) (0,5) (0,3) (0,3) (0,3) (0,4) (0,3) (0,4) (0,3) (0,4) (0,4) (0,4) (0,4) (0,4) (0,3) (0,3)}

    \end{axis}
    \end{tikzpicture}
    \vspace{-8pt}
    \caption{Distribution of number of changes in strictly-cloned patches.}
    \label{fig:distribution}
\end{minipage}}
\hspace{2pt}
\adjustbox{valign=t}{\begin{minipage}[r][][c]{0.245\textwidth}%
    \begin{tikzpicture}
    \begin{axis}[
        boxplot/draw direction=x,
        xmin=0.5,
        ymin=0.5,
        ymax=5.5,
        width=\linewidth,
        height=3.8cm,
        axis x line* = bottom,
        axis y line* = left,
        xlabel= \# files,
        boxplot={
            box extend=0.5
        },
        y dir=reverse,
        boxplot/draw/median/.code={%
          \draw[mark size=2pt,/pgfplots/boxplot/every median/.try]
            \pgfextra
            \pgftransformshift{
              \pgfplotsboxplotpointabbox
                {\pgfplotsboxplotvalue{median}}
                {0.5}
            }
            \pgfsetfillcolor{white}
            \pgfuseplotmark{*}
            \endpgfextra
          ;
        },
        xmajorgrids = true,
        xtick style={draw=none}, 
        ytick style={draw=none}, 
        ytick={0.5,...,5.5},
        xtick={1,3,5},
        y tick label as interval,
        yticklabels={Type A,Type B,Type C,Type D,Type E}
    ]
    
\boxplott{0}{1.5104166666666667}{1.0}{1.0}{2.0}{3.0}{1.0}{ (0,4) (0,4) (0,6) (0,6) (0,4) (0,4) (0,6) (0,5) (0,6) (0,6) (0,5) (0,5) (0,4)}
\boxplott{0}{1.3869969040247678}{1.0}{1.0}{2.0}{3.0}{1.0}{ (0,4) (0,6) (0,5) (0,5) (0,5) (0,5) (0,4) (0,6) (0,6) (0,4) (0,4)}
\boxplott{0}{1.3181818181818181}{1.0}{1.0}{1.25}{1.25}{1.0}{ (0,4) (0,3) (0,2) (0,2) (0,2) (0,2) (0,2) (0,2) (0,2) (0,2) (0,2)}
\boxplott{0}{1.3032786885245902}{1.0}{1.0}{1.0}{1.0}{1.0}{ (0,2) (0,2) (0,2) (0,2) (0,2) (0,2) (0,2) (0,2) (0,6) (0,2) (0,2) (0,2) (0,2) (0,2) (0,2) (0,4) (0,6) (0,2) (0,2) (0,4) (0,2) (0,2) (0,2) (0,2) (0,2)}
\boxplott{0}{1.39375}{1.0}{1.0}{2.0}{3.0}{1.0}{ (0,4)}
    
    \end{axis}
    \end{tikzpicture}
    \vspace{-6.4pt}
    \caption{Distribution of number of changed files in strictly-cloned patches.}
    \label{fig:distribution_file}
\end{minipage}}%
\vspace{-10pt}%
\end{figure}

Finally, \autoref{tab:clone_patterns} presents the occurrence of the SStuB (simple stupid bugs) patterns \cite{Karampatsis2020ManySStuBs4J} in change clones.
For instance, there are \numberToBeChecked{\np{200}} change clones that are ``Change Identifier Used'' instances in the ManySStuBs4J dataset. 
Those changes represent \numberToBeChecked{\percentage[e]{200}{5460}} of all change clones spotted by us, and \numberToBeChecked{\percentage[e]{200}{1851}} of all changes annotated with ``Change Identifier Used'' in the \nbAnalyzedPatchesToPrint analyzed multi-hunk patches.
The most frequent SStuB patterns in change clones are ``Wrong Function Name'' and ``Change Modifier''. We observed that most changes considered as change clones are not classified with any of the SStuB patterns, since the percentages shown in the second column of the table are far to sum up to 100\%. In fact, only \numberToBeChecked{\percentage[p]{1604}{5460}} change clones are annotated with one of the \numberToBeChecked{\np{16}} SStuBs.
Finally, with the numbers of the last column, we observed that ``Missing Throws Exception'', ``Change Unary Operator'', and ``More Specific If'' changes are more likely to be cloned, considering their total occurrences in multi-hunk patches.

\begin{table}[t]
    \centering
    \footnotesize
    \caption{SStuB patterns in change clones.}
    \label{tab:clone_patterns}
    \begin{tabular}{@{}l rr@{}}
\toprule
\multirow{3}{*}{SStuB} & Change clones & \% Change clones \\
{} & (total: \np{5460}) & in SStuB changes \\
\midrule
Change Identifier Used & \percentage[p]{200}{5460} & \percentage[e]{200}{1851} \\
Change Numeric Literal & \percentage[p]{13}{5460} & \percentage[e]{13}{393} \\
Change Modifier & \percentage[p]{318}{5460} & \percentage[e]{318}{420} \\
Change Boolean Literal & \percentage[p]{0}{5460} & \percentage[e]{0}{70} \\
\midrule
Wrong Function Name & \percentage[p]{479}{5460} & \percentage[e]{479}{1024} \\
Same Function More Args & \percentage[p]{186}{5460} & \percentage[e]{186}{288} \\
Same Function Less Args & \percentage[p]{65}{5460} & \percentage[e]{65}{100} \\
Same Function Wrong Caller & \percentage[p]{40}{5460} & \percentage[e]{40}{165} \\
Same Function Swap Args & \percentage[p]{56}{5460} & \percentage[e]{56}{81} \\
\midrule
Change Binary Operator & \percentage[p]{69}{5460} & \percentage[e]{69}{93} \\
Change Unary Operator & \percentage[p]{45}{5460} & \percentage[e]{45}{53} \\
Change Operand & \percentage[p]{18}{5460} & \percentage[e]{18}{80} \\
\midrule
Less Specific If & \percentage[p]{32}{5460} & \percentage[e]{32}{150} \\
More Specific If & \percentage[p]{49}{5460} & \percentage[e]{49}{59} \\
Missing Throws Exception & \percentage[p]{22}{5460} & \percentage[e]{22}{24} \\
Delete Throws Exception & \percentage[p]{12}{5460} & \percentage[e]{12}{19} \\
\bottomrule
    \end{tabular}
\end{table}

\section{Discussion}

\subsection{Implications}

In this study, we found out that \numberToBeChecked{\percentage[d]{\nbPatchesWithClonesNum}{\nbAnalyzedPatches}} of the multi-hunk patches contain change clones. This percentage is high, which indicates the need of repair tools that can produce change clones, i.e., that can apply identical or similar changes in different locations of the program.

Moreover, \numberToBeChecked{\percentage[d]{\nbSingleCloneChangePatches}{\nbPatchesWithClonesNum}} of these patches are strictly-cloned patches. This means that a repair tool could generate these patches by only applying the same or similar changes in the same or similar contexts, and no other change.

Finally, change clones are mostly of Type A, B, and E. Those change clone types are about identical changes being applied in different locations of the program, regardless of their contexts. This encourages automation for multi-hunk repair by only applying the same change in different locations. By only targeting Type A patches, the likely simplest case to automatize because the same change is applied in identical contexts, a system would fix \numberToBeChecked{\percentage[d]{480}{\nbPatchesWithClonesNum}} of the patches containing clones, which is \numberToBeChecked{\percentage[d]{480}{\nbAnalyzedPatches}} of all multi-hunk analyzed patches.

\subsection{Threats to validity}

There are \numberToBeChecked{three} main threats to the validity of our study, which are described as follows.

\vspace{5pt}
\noindent\textit{Manual analysis.} Our work relies on extensive manual analysis of patches. As in any manual work, we might have made mistakes when detecting change clones and classifying them. The manual analysis was performed by the two of us. At the beginning of the process, we analyzed \numberToBeChecked{\np{50}} patches together, discussed, and annotated them. This was a way to minimize the subjectivity in the task.

\vspace{5pt}
\noindent\textit{Bug fixing changes.} ManySStuBs4J is supposed to contain changes related to bug fixes. The authors of that dataset spotted and removed some recurring changes that are related to refactoring. However, it is not guaranteed that the dataset is \np[\%]{100} composed by bug fix changes. This is a threat to the validity of our study because we assume that the changes are bug fixing.

\vspace{5pt}
\noindent\textit{Single statement changes.} Our analysis was performed on single statement changes. Our findings cannot be generalized to patches containing blocks of changes.

\section{Related work}

\vspace{5pt}
\noindent\textit{Concepts.} The term \textit{change clone} used in this paper is related to \textit{code clone}, \textit{similarity preserving co-change (SPCO)} \cite{Mandal2014}, and \textit{systematic edit}. We already discussed the differences between change clone and code clone in \autoref{sec:definitions}. SPCO refers to clone fragments that co-changed and their similarity was preserved. This is similar but slightly different from change clone, because a pair of change clones is not necessarily a pair of code clones before the change was applied, e.g., Type E change clones. \textit{Systematic edits} are similar, but not identical, changes to many locations in the source code \cite{Meng2013}. On the contrary, change clones can be identical.

\vspace{5pt}
\noindent\textit{Clones and bugs.}
There are studies where clones and bugs were studied together, which are related works to ours.
Steidl and G\"{o}de \cite{Steidl2013}, for instance, investigated which clone features are relevant to predict incompletely fixed bugs.
Mondal et al. \cite{Mondal2019} investigated if the creation of code clones propagates temporarily hidden bugs from one code fragment to another.

\vspace{5pt}
\noindent\textit{Studies on patch analysis for program repair.}
In the context of automatic program repair, Sobreira et al. \cite{Sobreira2018defects4jdissection} performed a manual analysis of the Defects4J patches \cite{Just2014Defects4J} for discovering repair actions and patterns. Among other features, they found out a pattern in the patches whose name is ``copy/paste''. Their study is related to ours, but our study has a much larger scale in number of patches and we only focus on change clones and their \numberToBeChecked{five} different types.

\section{Final remarks}

In this paper, we reported on a study of change clones within multi-hunk patches written by developers. Our findings can guide researchers in improving the state-of-the-art of automatic program repair.

\vspace{5pt}
\noindent\textit{Future work.}
There are several opportunities for future work.
First, an automated solution for analyzing patches, such as PPD \cite{Madeiral2018}, could be created for detecting the change clone types, which would allow us to scale up our study. The patches annotated in this study could serve as ground-truth for evaluating that automated detection.
Second, we noted that the existing SStuB patterns are not enough to categorize the changes involved in clones. A future work would be to create a broader set of SStuB patterns so that at least most changes can be classified.
Third, for change clone types with identical changes (Types A, B, and E), it would be interesting to investigate how universal the change was in the source code of the analyzed project. For example, considering \autoref{code:typeB}, one might wonder if all instances of the call to \texttt{Long.valueOf} is deleted, or just some of them based on the context.
Fourth, since our study shows that the majority of change clones are Type B clones, which are identical changes with similar contexts, it would be beneficial to have a more fine-grained analysis on how different the contexts are across those identical changes. This would shed light on the possible challenges and solutions in developing automatic program repair that could generalize to different program contexts.
Finally, further studies could analyze change clones in multiple statement changes instead of single statement changes.

\section*{Acknowledgment}

We thank the anonymous reviewers and Martin Monperrus for their feedback on this paper.
This work was partially supported by the Swedish Foundation for Strategic Research under the TrustFull project, and by the Knut and Alice Wallenberg Foundation under the Wallenberg AI, Autonomous Systems, and Software Program (WASP).

\balance
\bibliographystyle{IEEEtran}
\bibliography{references}

\end{document}